\documentclass[aps,pra,twocolumn,amsmath,amssymb,showpacs,fixfloats]{revtex4}
\newcommand{\be}{\begin{equation}}
\newcommand{\ee}{\end{equation}}
\newcommand{\bea}{\begin{eqnarray}}
\newcommand{\eea}{\end{eqnarray}}
\newcommand{\ud}{\mathrm{d}}

\usepackage{units}
\usepackage{graphicx}
\usepackage{amsmath}
\usepackage{epsfig}

\begin{document}
\title{Influence of the energy-band structure on ultracold reactive processes
in lattices}

\author{H. Terrier, J.-M. Launay, and A. Simoni}
\affiliation{
Institut de Physique de Rennes,  UMR CNRS 6251, Universit\'e de
Rennes 1, F-35042 Rennes, France
}


\begin{abstract}

We study theoretically ultracold collisions in quasi one-dimensional optical traps
for bosonic and fermionic reactive molecules in the presence of a periodic 
potential along the trap axis. Elastic, reactive, and {\it umklapp} processes
due to non-conservation of the center of mass motion are investigated
for parameters of relevant experimental interest. The model naturally
keeps into account the effect of excited energy bands and is particularly
suited for being adapted to rigorous close-coupled calculations.
Our formalism shows that a correct derivation of the 
parameters in tight-binding effective models must include the strong momentum
dependence of the coupling constant we predict even for deep lattices. 

\end{abstract}

\maketitle


\section{Introduction}

Ultracold gases confined in optical lattices represent an extremely
active area of theoretical and experimental research for quantum few and
many-body physics \cite{2005-IB-NP-23,2008-IB-RMP-885}. The flexibility
in the choice of the lattice parameters and the variety of trappable
atomic and molecular species allow the properties of these systems to
be controlled with unprecedented accuracy. A variety of phases has been
predicted and several have been observed in celebrated experiments on
{\it atomic} quantum gases trapped in lattices of various dimensionality
and structure \cite{Greiner2002a,2004-BP-NAT-277,2006-ZH-NAT-1118}.
Over the last decade, different dynamical aspects such as
transport, few-body correlations, and the occurrence of geometric
resonances have been studied experimentally under lattice confinement
\cite{2004-HO-PRL-160601,2004-BLT-PRL-190401,2005-HM-PRL-210401,2005-CDF-PRL-120403,2010-PRL-EH-153203}.
It is now also possible to prepare cold or even Bose-condensed gases in
excited bands of optical potentials \cite{2011-GW-NP-147}.

Understanding the microscopic few-body collisional
interaction is essential to model the macroscopic behavior
of a gaz held in a lattice. Several papers focused
on the two-atom problem in conditions where an harmonic
confining potential restricts the motion to one or two dimensions
\cite{1998-MO-PRL-938,Petrov2000a,2002-ELB-PRA-013403,AM-2010-PRL-073202,2012-SS-PRL-073201}.
The recent production of ultracold {\it molecular} samples has opened
the way to the study of reactive processes at extremely low temperatures
both in free space and in confined geometries. The combined effect of
confinement and of a polarizing electric field has been studied both
theoretically and experimentally in \cite{2011-MHGDM-NP-502}, where the
authors demonstrate that repulsion between polarized molecules held in
a quasi-2D pancake geometry strongly suppresses the reaction probability
and stabilizes the gaz. A theoretical analysis of reactions has also been
carried out in quasi-1D optical tubes \cite{2015-AS-NJP-013020}. In all
these studies hopping between lattice sites is not included and the
potential in the effectively free dimensions remains flat.

The periodic nature of the trapping optical lattice has been taken
explicitly into account within tight binding lattice models 
\cite{2004-OPF-PRL-080401,2008-NN-PRA-023617} and using a more general
formalism \cite{2006-MW-PRL-012707,2010-HPB-PRL-090402}. Little is
known about reaction dynamics in a periodic potential.  Experimental work
carried out at JILA on reactive polar molecules addresses the effect
of a weak lattice (a corrugation) superimposed along the axis of an
array of tubes \cite{2012-AC-PRL-080405}. The authors speculate on the
origin of the observed reaction rates and show that one can interpret
the observed suppression of such inelastic processes as a manifestation
of the quantum Zeno effect \cite{2014-BZ-PRL-070404}.

Main goal of the present paper is to model the two-body {\it reaction}
dynamics in 1D geometries similar to the aforementioned experiment
\cite{2012-AC-PRL-080405} using a more direct collisional approach. We
solve an effective model where the short range reaction dynamics
is represented by a completely adsorbing boundary condition. The
key novelty of our formalism is the definition of a set of reference
wavefunctions that include in a rigorous way the effect of the excited
energy bands. While the formalism is applied to a simple yet realistic
purely 1D model with a point-like interaction, it can in principle be readily
incorporated in a full 3D close-coupled calculation such as the one of
Ref.~\cite{2015-AS-NJP-013020}.

The paper is organized as follows. Sec.~\ref{sec2} describes the model,
defines the scattering observables in terms of two-body Bloch functions,
and introduces our numerical approach.
Sec.~\ref{sec3} presents results for lattices of different strength for
reactive molecules of both bosonic and fermionic nature. Approximations
are developed and discussed. A short conclusion summarizes
and puts into perspective the present work.

\section{Theoretical model}
\label{sec2}

We consider ultracold molecules in a strongly confining harmonic potential
along radial directions $x$ and $y$ in the presence of a weaker
periodic sinusoidal optical potential along the axial direction $z$.
The harmonic trapping frequency $\omega_\perp$ is the same for
both radial directions. In the absence of the optical lattice, if the
interaction potential is sufficiently short ranged one can describe
the system as quasi-one-dimensional along the longitudinal $z$ axis.
The practical criterion for van der Waals interactions with dispersion
coefficient $C_6$ and particles of mass $m$ and reduced mass $\mu=m/2$ 
is that the van der Waals
length ${\bar a}=0.5 (2 \mu  C_6/ \hbar^2)^{1/4}$ be much smaller than the
transverse oscillator length $a_{\perp}=\sqrt{\hbar/ \mu \omega_\perp}$.

The K-Rb fermionic molecules studied at JILA 
have been found experimentally to be highly reactive~\cite{2010-SO-SCI-853}.
In order to address this experimentally relevant case,
this work will focus on universal reactive processes,
in which all particle flux reaching a suitable short range region gives rise to
reaction with unit probability \cite{AM-2010-PRL-073202}. 
The typical size of this region can be taken to be on
the order of $\sim 40 a_0$, such that the potential beyond this boundary
is well approximated by an isotropic van der Waals potential. Under
such conditions the short-range collision dynamics for {\it
bosonic} particles can be summarized by a pseudopotential $U_a=g
\delta(z_2-z_1)$ with complex coupling constant. The coupling constant
can in turn be simply expressed in terms of the geometric parameters
of the trap and of the van der Waals length as $g=2 \hbar
\omega_\perp {\bar a} (1-i)$ \cite{AM-2010-PRL-073202}.

It is important for the following to stress that the pseudopotential
conserves real CM momentum (as opposed to the quasimomentum) of the
system. Also note that the pseudopotential cannot be directly used to
model Fermions since due to Pauli exclusion principle the wavefunction
strictly vanishes at zero interparticle separation. However, as it will
be detailed in Sec.~III D one can take advantage of a boson-fermion
mapping procedure in order to apply with little modification the current
formalism to fermionic particles as well.

Finally, the optical lattice along the longitudinal direction is taken
of the form $U_{\rm L}(z)=u \sin^2(k z) = u \left[ 1-\cos({k_{\rm L}
z})\right] /2$, where $u$ is the lattice depth, $k$ the laser wavevector,
$k_{\rm L} \equiv 2k$ the lattice wavevector, related to the lattice
period $a=2\pi/k_{\rm L}$. We assume that this lattice does not modify
the effective 1D interatomic potential. This is only justified when the
harmonic oscillator frequency $\omega_{\rm L} = k_{\rm L} \sqrt{u/2m}$,
obtained by Taylor expanding $U_{\rm L}$ near a minimum of the lattice
potential, is much smaller than $\omega_\perp$. The lattice depth will
henceforth be expressed in terms of the {\it lattice} recoil energy for
one molecule defined as $E_{\rm R}=\hbar^2 k_{\rm L}^2/(2m)$. Note this
quantity is four times larger than the {\it laser} recoil energy $\hbar^2
k^2/(2m)$ which is often used in experimental work.

\subsection{Band structure and free Bloch waves}

In the absence of the effective interatomic potential the wavefunction
of a pair of molecules is described by a product of 1D Bloch waves $
| \phi_{p_1}^m \rangle  \otimes |\phi_{p_2}^n \rangle$, labeled by
quasi-momenta $p_\alpha$ for particle $\alpha$ in energy zones $m$
and $n$, respectively.  The energy of a single-particle Bloch state
$|\phi_{p}^m \rangle$ is ${\cal E}_m(p)$ and the single-atom states
are quasi-momentum normalized $\langle \phi_{p^\prime}^m| \phi_p^n
\rangle=\delta(p'-p) \delta_{mn}$.  For describing the effects
of the interaction it is convenient to introduce the relative and CM
quasi-momenta $\{ q=(p_1-p_2)/2,Q=p_1+p_2 \}$ and the two-particle
state $|\Phi_{qQ}^{mn} \rangle = |v_{mn}(q,Q)|^{-1/2} \left( |\phi_{Q/2
+q}^m \rangle \otimes | \phi_{Q/2 - q}^n \rangle \right)$. Such state
has energy $E_{mn}(q,Q)={\cal E}_m(Q/2 +q)+{\cal E}_n(Q/2 -q)$ and
$v_{mn}(q,Q)={\partial E_{mn}}/{\partial (\hbar q)}$ is the relative
{\it group} velocity for particles in zones $m$ and $n$.  In this way
pair states are quasi-momentum normalized for the CM quasi-momentum
and energy or flux normalized for the relative quasi-momentum, {\it i.e.}
$\langle \Phi_{q' Q'}^{m'n'} | \Phi_{q Q}^{mn} \rangle=\delta\left[
E_{mn}(q',Q')-E_{mn}(q,Q)\right] \delta(Q'-Q) \delta_{mm'} \delta_{nn'}$.
For notational simplicity, when possible we will henceforth condense
the double zone index into a single Greek letter $\alpha \equiv (m,n)$.

When two molecules collide with relative quasi-momentum $q$ and CM
quasi-momentum $Q$ the interatomic potential can induce coupling to
different states $\Phi_{q' Q'}$ such that total energy is conserved
$E^\prime=E$ and the CM quasimomentum varies by a multiple of a lattice
vector $Q^\prime=Q \pmod {k_{\rm L}}$.  Such processes are known as {\it
umklapp} collisions in solid state physics (see e.g. Ref~\cite{Ashcroft})
and have been experimentally observed in \cite{2006-GKC-PRL-020406} where they were
described in terms of a phase matching condition. In a quasi-1D lattice such
condition of energy and momentum conservation is restrictive and only
a small number of allowed states exist for given $E$ and $Q$. {\it Umklapp}
processes are more conveniently discussed by choosing the fundamental reciprocal
lattice cell $\{ q , Q \} \in [-k_{\rm L}/2,k_{\rm L}/2] \times [0,
k_{\rm L} ]$ where $Q$ is unambiguously defined. Thus, when restricted to
this specific unit cell {\it umklapp} collisions strictly, not only $\pmod {k_{\rm L}}$, 
do conserve $Q$.

\begin{figure}[hb!]
\centering
\includegraphics[width=0.9\columnwidth]{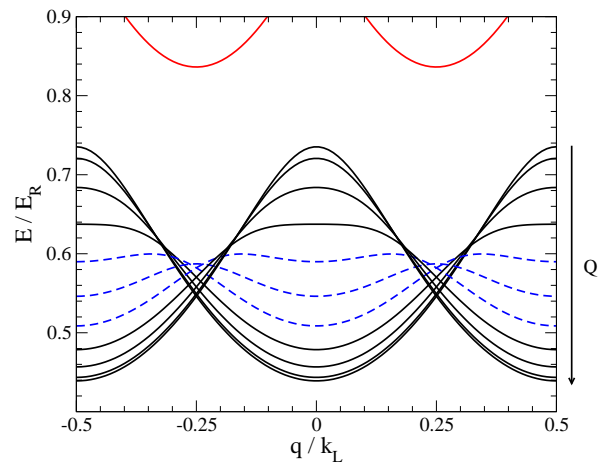}
\caption{
Two-particle energy as a function of $q$ in an optical
lattice of depth $u=0.5 ~ E_{\rm R}$ for discrete values of total
quasimomentum $Q$ varying from $0$ to $k_{\rm L}$ in step of $0.1~  k_{\rm
L}$ in the sense of the arrow. 
Both molecules are in the fundamental
band. The dashed curves correspond to CM quasimomenta for which {\it umklapp}
collisions are allowed (see text). The lower limit of the first excited
energy band is visible in the upper part of the panel.
}
\label{fig1} 
\end{figure}

Fig.~(\ref{fig1}) shows the relative-motion energy dispersion relation of two
molecules in the fundamental band for selected values
of $Q$. It is easy to show that square lattice periodicity implies that the reciprocal
lattice points $(q,Q)$ and $(q-k_{\rm L},Q)$ only differ by a reciprocal lattice vectors.
The energy is thus periodic in $q$ at the cell edges, {\it i.e.}
$E_{mm}(-k_{\rm L}/2,Q)=E_{mm}(k_{\rm L}/2,Q)$ for particles in the same band. 
If particles belong to different energy bands the former symmetry relation
reads $E_{mn}(-k_{\rm L}/2,Q)=E_{nm}(k_{\rm L}/2,Q)$.  

We consider
in this work identical particle scattering and by the symmetrization
principle the range of $q$ can be restricted to half space only ($q\geq
0$) if particles are in the same energy band. For particles in
different bands, one has to consider either $q \geq 0$ and both $(m,n)$
and $(n,m)$ combinations, or both positive and negative values of $q$
with the restriction $m>n$. With this proviso, if for a given $Q$
and energy $E$ of the incoming molecules the dispersion relation
$E_{mn}(q,Q)$ presents at least a maximum or minimum in $q$, {\it umklapp}
collisions in the given band will occur at energies $E$ such that the equation $E(q,Q)=E$
admits a double solution (with the restriction $q>0$ if $m=n$). Sample values of $Q$ for which this condition
is verified are shown by dashed curves in the figure. On the converse,
if the dispersion relation is monotonic no {\it umklapp} processes will be
possible. Note that if the lattice is weak or collisions take place in
excited bands the energy gap will become sufficiently small that
{\it umklapp} processes can take place not only intra-, as in the case of
Fig.~(\ref{fig1}), but also inter-band.

If the lattice is strong enough the dispersion relation is explicitly known
in the tight binding approximation. For one molecule in energy band $n$
it can be written as ${\cal E}_m(p_1)= D_m - J_m 
\cos(a p_1)]$ \cite{Ashcroft} where $|J_m|$ is the energy half-bandwidth
and the constant $D_m$ in the harmonic approximation is related to
the harmonic frequency at the bottom of the well as $D_m= \hbar \omega_{\rm L}
(m+1/2)$. Since potential is separable, energy is additive $E_{mn} = {\cal
E}_m(p_1) + {\cal E}_n(p_2)$ and simple algebra gives 
\bea
\label{Enm}
E_{mn}(q,Q)&=& D_m + D_n +
 (J_m - J_n) \sin(a Q/2) \sin(a q) \nonumber \\
&-& (J_m + J_n)  \cos(a Q/2) \cos(a q) .
\eea
It is easy so show that $E_{mn}(q,Q)$ is monotonic in the interval $q \in [
0 , k_{\rm L} /2 ]$ for any $Q$ if $n=m$, otherwise it admits therein a single maximum
or minimum for $Q \neq 0$. The {\it umklapp} collisions
will therefore occur in the tight binding limit if and only if the
colliding particles belong to different single-particle energy bands.


\subsection{Scattering wavefunction and scattering observables}

The incoming wavefunction for the collision will be taken as a Bloch wave $\Phi_{q,Q}^\alpha(z,Z)$ with
relative group velocity $v<0$, {\it i.e.} moving opposite to the $z$ direction.
Outside the range of the interatomic potential ($z \to \infty$) the wave function is written
as the sum of an incoming plus outgoing scattered Bloch waves
\be
\label{asyfun}
\Psi_{qQ}^{\alpha}= \Phi_{qQ}^{\alpha} \pm \sum_{q^\prime Q^\prime \alpha'} \Phi^{\alpha'}_{q^\prime Q^\prime}  
S(q^\prime Q^\prime \alpha' \gets q Q \alpha)
\ee
where the upper (lower) sign holds for bosons (fermions). The sum
includes all purely real relative quasi-momenta $q^\prime$ such that
$E_{\alpha'}(q^\prime,Q^\prime)=E_{\alpha}(q,Q)$ in zone $\alpha'$ for
$Q^\prime=Q \pmod {k_{\rm L}}$.  
Since we are imposing outgoing boundary conditions,
the scattered waves on the right hand side must be chosen with relative
group velocity in the direction of the positive $z$ axis for $z>0$.

The sum also contains functions that decrease exponentially with increasing
$z$ characterized by complex relative quasi-momentum with $\Im q >0$.
In the language of scattering theory, we will often term the propagating
(evanescent) waves with $\Im q = 0$ ( $\Im q \neq 0$) the open (closed)
channels of the collision. An algebraic procedure to determine
both kind of Bloch waves will be the subject of Sec.~III C.

For a real non-absorbing potential, the choice of flux normalized reference functions
and conservation of probability result in unitarity of the $S$ matrix
block formed by all the $S(q^\prime Q^\prime \alpha' \gets q Q \alpha)$ elements with
purely real $q$ and $q^\prime$. This is not any longer true for our potential that aims
at modelling reactive processes through the introduction of a nonzero
imaginary part. It is indeed the difference to unitarity that gives the
reaction probability; See Eq.~\eqref{preact} below.

The scattering matrix encompasses all information about the scattering
process. Our conventions on the propagation directions of the incoming
and scattered waves allow the same expressions for the scattering
observables to be used as in the no-lattice case \cite{2015-AS-NJP-013020}. Thus, the
elastic scattering rate is expressed in terms of the scattering matrix as
\be
\label{kel}
   {\cal K}^{\rm el}_\alpha(q,Q) = |v_\alpha(q,Q)|   \left|  1 - S(q Q \alpha \gets q Q  \alpha)  \right|^2 
\ee
Another quantity of experimental interest is the reaction probability
\be
 \label{preact}
     P^{\rm r}_\alpha(q,Q)=1-\sum_{q^\prime Q^\prime \alpha'} |S(q^\prime Q^\prime \alpha' \gets q Q \alpha)|^2
\ee
which in the formalism stems from the lack of unitarity of the
$S$-matrix due to the complex nature of the coupling constant. The summed
probability for {\it umklapp} processes, that can be interpreted as superelastic
collisions taking place at fixed {\it total} energy, is given by
\be
   P^{\rm u}_\alpha(q,Q)  =  \sideset{}{'} \sum_{q^\prime Q^\prime \alpha'} |S(q^\prime Q^\prime \alpha' \gets q Q \alpha)|^2
\ee
where the sum $\sum^\prime$ is restricted to  $\{q^\prime Q^\prime \alpha' \}
\neq \{q Q \alpha \}$. Such scattering quantities are in principle measurable
in a reference frame in uniform motion with velocity equal to the CM
group velocity $v_{\rm CM}={\partial E}/ {\partial (\hbar Q)}$ of the colliding particles.

The effect of the lattice can be summarized in quasimomentum-dependent scattering 
lengths defined for bosons and fermions respectively as 
\be
    a^{\rm B}_\alpha(q,Q) = \frac{\hbar}{\mu |v_\alpha(q,Q)| k_\alpha(q,Q)}
\ee
and
\be
a^{\rm F}_\alpha(q,Q) = -\frac{\hbar \, k_\alpha(q,Q)}{\mu |v_\alpha(q,Q)|}
\ee
where $\mu=m/2$ is the reduced mass and $k_\alpha(q,Q) \equiv i\left[1-S(q
Q \alpha \gets q Q \alpha)\right]/\left[ 1+S(q Q \alpha \gets
q Q \alpha)\right]$. If one writes as usual $S=e^{2 i \delta}$,
the quantity $k$ is seen to simply represents the tangent of the
(generally complex) phase shift $\delta$. Note that the $a^{\rm F,B}$
above reduce to the standard ones in free space in the absence of the
lattice. 
Finally, boson scattering can
also be conveniently described by introducing an effective
momentum-dependent coupling constant in band $\alpha$
\be
 \label{geff}
   g^{\rm eff}_\alpha(q,Q) = - \frac{ \hbar^2}{\mu \, a^{\rm B}_\alpha(q,Q)}  .
\ee
This quantity is the effective interaction that after averaging over
a period of the center-of-mass coordinate gives a scattering matrix element in the
elastic channel equal to the actual one. The imaginary part of $g_{\rm
eff}$ accounts for the combined effect of reactive and {\it umklapp} processes.
In the absence of optical lattice $g^{\rm eff}(q,Q)$ reduces to $g$ as expected.

\subsection{Determining the reference functions}
\label{drf}

For notational convenience in next two sections we adopt as the units of
momentum, length, and energy the lattice vector $k_{\rm L}$, its inverse
$k_{\rm L}^{-1}$ and the lattice recoil energy $E_{\rm R}$, respectively. 
The coupling constant will thus be expressed in units $E_{\rm R}/k_{\rm L}$.

The reference functions used in Eq.~\eqref{asyfun} can be built as follows.
According to the Bloch theorem, even in the presence of the interaction
the scattering wavefunction transforms as a Bloch function $\Psi_{q Q}(z,Z+ 2\pi )= e^{i 2 \pi Q  } \Psi_{q,
Q}(z,Z)$ under CM coordinate translations of period $2 \pi$.
Bloch's theorem allows then $\Psi$ to be expressed in the general form
$\Psi_{qQ}(z,Z) = e^{iQZ} \sum_n e^{in Z} \psi_{qQ,n}(z)$.

In order to represent the interacting $\Psi$, non interacting reference functions must therefore
be built with the given (real) value of $Q$.
To this aim, Bloch theorem now applied to both the CM and relative
coordinates implies that the pair wavefunction in the absence of the interaction
can be written in the form $\Phi_{q, Q}(z,Z)= e^{i (Q Z + qz)} u_{qQ}(z,Z)$ with
$u_{qQ}$ periodic with the lattice periodicity. Similarly, it can be
easily verified by direct differentiation that any spatial derivative
of $\Phi_{q Q}$ can be expressed in Bloch's form as well. In particular,
$\chi_{qQ}(z,Z) \equiv \partial_z \Phi_{q Q}(z,Z)= e^{i (Q Z + qz)}
v_{qQ}(z,Z)$ with $v_{qQ}= iq u_{qQ} + \partial_z u_{qQ}$ a function periodic on
the lattice.

With these definitions, the second order time independent Schr{\"o}dinger equation can be rewritten as a set of
two coupled partial differential equations of order one 
\bea
\label{eqqr}
 i \frac{\partial}{\partial z} u_{qQ} -i v_{qQ} = q u_{qQ} \nonumber \\
 i \frac{\partial}{\partial z} v_{qQ} + \dfrac{i}{2} \left[ E - \frac{1}{2} (-i \frac{\partial}{\partial Z} 
+Q)^2 - V_{\rm L}(z,Z) \right] u_{qQ} 
= q v_{qQ}  
\nonumber \\
\eea
in the form of an eigenvalue problem for the unknown spatial functions
$u_{qQ}$ and $v_{qQ}$ and the eigenvalue $q$. The operator on the lhs
can be considered a matrix differential operator acting on column
vector wavefunctions ${\vec w}_{qQ} \equiv ( u_{qQ}, v_{qQ} )^t$ in
a lattice unit cell.  One can verify that such operator is hermitian
with respect to the symplectic inner product $\langle {\vec w}_{q'Q}
| {\vec w}_{qQ} \rangle_s \equiv \langle u_{q'Q} | v_{qQ} \rangle -
\langle v_{q'Q} |  u_{qQ} \rangle$, where the scalar product on the rhs
is the standard one over a unit cell. Since the symplectic inner product
is not positive-definite it is not possible to show that $q$ is real
in general. One can however deduce following the standard proof
that if $q' \neq q^*$ then the symplectic orthogonality condition $\langle
{\vec w}_{q'Q} | {\vec w}_{qQ} \rangle_s \propto \delta_{q' q^*}$ holds.

Additional properties follow from the lattice symmetries. First,
since the lattice potential $V_{\rm L}$ is real, taking the conjugate
of \eqref{eqqr} shows that if ${\vec w}_{qQ}$ is eigenfunction with
eigenvalue $q$ then ${\vec w}_{qQ}^*={\vec w}_{-q^*-Q}$ is eigenfunction
of the corresponding problem with quasimomentum $-Q$ and eigenvalue
$q^*$ (time-reversal property). Moreover, the particles being identical,
irrespective of the nature of $V_{\rm L}$ one has $V_{\rm L}(-z,Z)=V_{\rm
L}(z,Z)$. Application of the permutation operation $z\to -z$ to
\eqref{eqqr} shows that if $q$ is eigenvalue with eigenfunction $( u_{qQ},
v_{qQ} )$ then $-q$ is also eigenvalue with eigenvector $( u_{-qQ}(z,Z),
v_{-qQ}(z,Z) ) = ( u_{qQ}(-z,Z), -v_{qQ}(-z,Z))$. Finally, reflection
of both coordinates $(z,Z)\to (-z,-Z)$ about the center of symmetry of
the lattice also leaves $V_{\rm L}$ invariant. One concludes that if $(
u_{qQ}, v_{qQ} )$ solves Eq.~\eqref{eqqr} with eigenvalue $q$, $( u_{-q
-Q}(z,Z), v_{-q -Q}(z,Z) ) = ( u_{qQ}(-z,-Z), -v_{qQ}(-z,-Z))$ will also
be 
solution of the equation with eigenvalue $-q$ and CM
quasimomentum $Q$.  Note that if $q$ is purely real or purely imaginary,
only two of the four solutions with quasimomenta $\pm q$ and $\pm q^*$
generated by the symmetry operations above will be linearly independent.

Equations~\eqref{eqqr} can be solved algebraically noting that the periodic
nature of the functions $u_{qQ}$ and $v_{qQ}$ allows the latter to be
developed in a 2D Fourier series
\be u_{qQ}(z,Z)=   \sum_{\vec K} a_{\vec K} (q,Q) e^{i {\vec K} \cdot {\vec r}} 
\ee
and 
\be
  v_{qQ}(z,Z)=  \sum_{\vec K} b_{\vec K} (q,Q) e^{i  {\vec K} \cdot {\vec r}}   ,
\ee
where the sum is over all reciprocal lattice vectors $\vec K$, and using
standard numerical eigenvalue solvers. The resulting families of discrete
eigenvalues $q$ depend on the two real parameter $Q$ and $E$. Since
the bidimensional quasimomentum vector of components $(q,Q)$ is defined $\pmod {\vec
K}$, as in the discussion of {\it umklapp} processes it is convenient to choose
the specific unit cell $( \Re q,  Q ) \in [-1/2,1/2] \times [0, 1 ]$.
Note that the imaginary part $\Im q$ is to be left unrestricted since
Bragg periodicity only concerns the real part of the quasimomentum.
\begin{figure}[h!]
\centering
 \includegraphics[width=0.9\columnwidth]{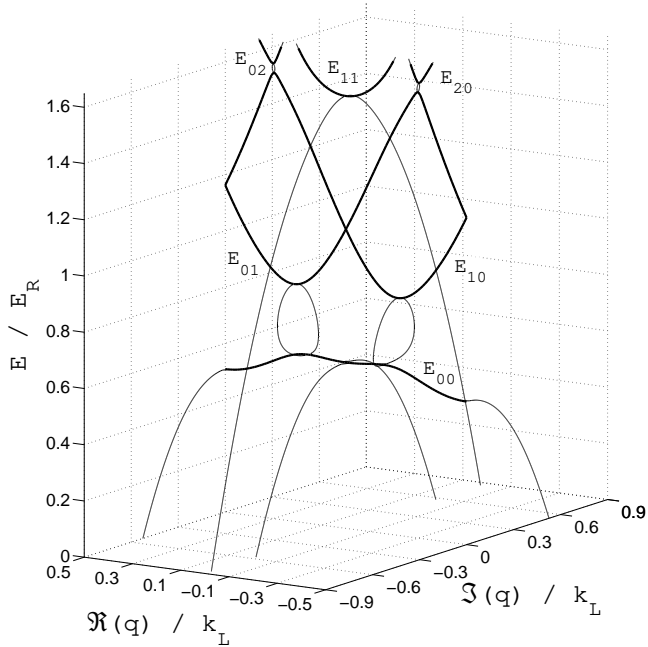}
\caption{
Evolution with $E$ of the complex relative quasimomentum determined from
the solution of Eq.~\eqref{eqqr} for $Q=0.6 ~k_{\rm L}$ and $u=0.5 ~E_{\rm R}$.
Propagating Bloch waves are characterized by a purely real $q$
(heavy lines) describing energy bands $E_{mn}$. Thin lines represent the quasimomenta of evanescent waves
with $\Im q \neq 0$ (see text). Two additional curves with larger $|
\Im q |$ and $\Re q=0 $ fall in the range of the figure but have not
been drawn to improve plot readability.
}
\label{figlines}
\end{figure}

It is instructive to study the evolution of eigenvalues $q$ with total
energy at fixed values of total quasimomentum; See Fig.~(\ref{vrelfig}) for
the selected case $Q=0.6$. With reference to the figure, in the allowed energy regions the energy band
structure highlighted by thick lines is retrieved as expected. One can
recognize in particular the fundamental band dispersion $E_{00}$ with two
extrema already depicted in Fig.~(\ref{fig1}) and the first excited band
structure formed by the $E_{01}$ and $E_{10}$ components. The symmetry
$E_{01}(q,Q)=E_{10}(-q,Q)$ is apparent. The bottoms of higher energy
bands $E_{11}$, $E_{20}$ and $E_{02}$ are also visible.

If $E$ is below the lowest limit of the fundamental band all $q$
have nonvanishing imaginary part and no propagating waves exist. In Fig.~(\ref{vrelfig})
one identifies four branches located in the $\Re q =0$ plane
and two in the $\Re q =\pm 1/2$ planes, respectively. That is, below the
minimum of $E_{00}$ one finds six intersections between any horizontal
plane and the curves of complex $q$ depicted in the plot. 

These curves correspond to evanescent-character waves and merge at
critical points with the dispersion curves of the authorized energy zones. In
fact, $|\Im q |$ decreases with increasing $E$, and as $|\Im q|  \to 0$
two branches merge with $E_{00}$ at the edge of the Brillouin zone $\Re
q = \pm k_{\rm L}/2$ where they become of propagating character. At
slightly larger energy two branches with purely imaginary argument $q$
merge with the curve $E_{00}$ at the relative minimum of the latter. The
two remaining curves with $\Re q =0$ in the figure merge at the bottom
of $E_{11}$ at still larger energies ($E \simeq 1.5$).

The behavior near the upper limit of $E_{00}$ is particularly interesting.
As the value of energy varies from below to above the upper band
limit two new branches with $\Im q \neq 0$ appear above each maximum.
These curves are non planar and join again near the minima at the bottom
of the $E_{10}$ and $E_{01}$ bands.  The existence of these ``smoke
rings'' connecting band edges is necessary since quantum states cannot
disappear as the parameter $E$ is made to vary. Similar ring structures
connecting the top of $E_{10}$ and $E_{01}$ with the bottom of $E_{20}$
and $E_{02}$ are also (barely) visible in the figure.

\subsection{Scattering matrix calculation}
 \label{smc}
Next we note that $2\pi$-periodicity in the $Z$ coordinate alone
allows one to write $\Phi_{qQ}(z,Z)=  e^{iQZ} \sum_n  e^{i n Z} \eta_{qQ,n}(z)$
and $\chi_{qQ}(z,Z)  =  e^{iQZ} \sum_n  e^{i n Z} \xi_{qQ,n}(z)$, where
\bea
\eta_{qQ,n}(z) &=& \frac{e^{iqz}}{2 \pi} \int_0^{2\pi} dZ e^{-i n Z} u_{qQ}(z,Z)  \nonumber \\ 
 \xi_{qQ,n}(z) &=& \frac{e^{iqz}}{2\pi}  \int_0^{2\pi} dZ e^{-i n Z} v_{qQ}(z,Z) .
\eea
For a given truncation order of the Fourier series in $Z$ it is possible
to arrange the coefficients $\eta_{qQ,n}$ and $\xi_{qQ,n}$ into
finite square solution matrices ${\boldsymbol \eta}_{\pm}$ and ${\boldsymbol
\xi}_{\pm}$, where ${\boldsymbol \eta}_{-}$ (resp. ${\boldsymbol
\eta}_{+}$) contains along the columns propagating Bloch waves with $v<0$
($v>0$) and evanescent Bloch waves exponentially decreasing
towards $z>0$ ($z<0$). 

Combining the symmetry properties stated below Eq.~\eqref{eqqr} one can easily prove the
symmetry relation
$\eta_{qQ,n}^*(z)=\eta_{-q^* Q,n}(z)$ and $\xi_{qQ,n}^*(z)=\xi_{-q^* Q,n}(z)$.
As already remarked, four independent degenerate solutions 
may exist with momenta $\pm q$ and $\pm q^*$. In this case, which only may occur for closed channels, we find it convenient 
to form
real reference functions according to $(\eta_{\pm qQ,n}+\eta_{\mp q^*Q,n})$ and 
$-i(\eta_{\pm qQ,n}-\eta_{\mp q^*Q,n})$.
Analogous definitions are used to construct the derivatives $\xi$.

With these conventions the matrix form of Eq.~\eqref{asyfun} is
\be
\label{asyfunmat}
{\boldsymbol \psi}(z) = {\boldsymbol \eta}_-(z) +  {\boldsymbol \eta}_+(z)  {\mathbf S}
\ee
where $\boldsymbol \psi$ has elements $\psi_{qQ,n}$,
while the $z$ derivative is expressed as
\be
\frac{\partial}{\partial z} {\boldsymbol \psi}(z) = {\boldsymbol \xi}_-(z) +  {\boldsymbol \xi}_+(z)  {\mathbf S} .
\ee

In order to determine the scattering matrix we note that our zero-range model
interaction conserves {\it real} (as opposed to quasi) total momentum
and imposes a discontinuity to the relative coordinate derivative at
the origin
\be
\label{condor}
 \left. \frac{\partial}{\partial z} \psi_{qQ,n} \right|_{0^+} -  \left. \frac{\partial}{\partial z} \psi_{qQ,n} \right|_{0^-} =  
2 \left.  \frac{\partial}{\partial z} \psi_{qQ,n} \right|_{0^+} = \frac{g}{2}  \psi_{qQ,n}(0) 
\ee 
where the first equality follows from the even character of the bosonic
wavefunction. This set of conditions can be summarized in the logarithmic
derivative of the matrix solution evaluated at the origin ${\mathbf
Z}\equiv (\partial_z {\boldsymbol \psi}) {\boldsymbol \psi}^{-1} =
(g/4) {\mathbf 1}$. Finally, imposing the asymptotic form \eqref{asyfunmat} and
following the standard matching procedure \cite{2011-AS-JPB-235201} the scattering matrix can
be determined in terms of $\mathbf Z$ by solving the linear system

\be
\label{zseq}
  \left[ {\boldsymbol \xi}_+(0^+)  - {\mathbf Z} {\boldsymbol \eta}_+(0   \right]
  {\mathbf S}  = \left[ {\mathbf Z} {\boldsymbol \eta}_-(0) - {\boldsymbol
  \xi}_-(0^+)     \right]
.
\ee

For fermions the condition dictated by the zero-range potential is of
no practical use since it is identically verified by any wavefunction
asymmetric in $z$. Use of more sophisticate pseudopotentials
valid for fermions can be avoided by the simpler prescription of
Ref.~\cite{2004-BEG-PRL-133202}. In this approach, resulting from a mapping between
bosons and fermions valid in dimension one, the fermionic problem is
solved as if the particles were bosons with a mapped coupling constant
\be
\label{map1}
\frac{g_{\rm map}}{a_\perp  \hbar \omega_\perp }= -\frac{ a_\perp}{a^{\rm F}_0}  .
\ee
Here the $a^{\rm F}_0$ is the 1D scattering length for fermions in the
absence of the optical lattice, which can be determined analytically for
universal scattering in tubes \cite{AM-2010-PRL-073202}. Substituting the latter 
equation into \eqref{map1} one obtains
\be
\label{map2}
      \frac{g_{\rm map}}{a_\perp \hbar \omega_\perp}= \frac{a_\perp^3}{12 {\bar a}^2{\bar a}_1} (1-i) ,
\ee
where ${\bar a}_1=1.064~\bar a$.
Note that fermions, which due to Pauli principle are weakly interacting, are
mapped to a bosonic problem implying strong interactions, {\it i.e.} the
$g_{\rm map}$ constant above is large in natural units. The scattering matrix
${\mathbf S}_{\rm map}$ is computed with the mapped interaction exactly
as for bosons and as a final step simply transformed to the physical
fermionic counterpart ${\mathbf S}^{\rm F}=-{\mathbf S}_{\rm map}$.

\subsection{Wigner threshold laws}

\begin{figure}[h!]
\centering
 \includegraphics[width=0.75\columnwidth]{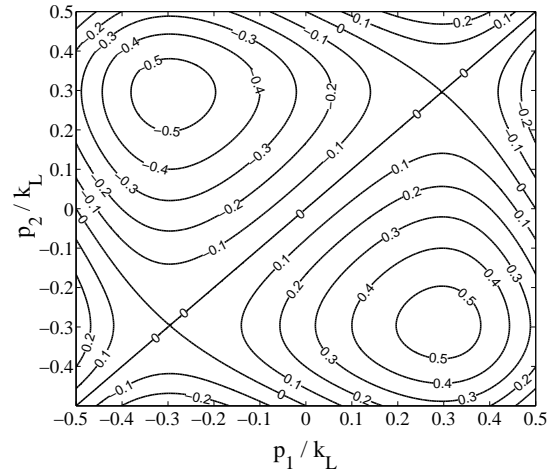}
\caption{
The relative group velocity $v$ of two particles in the first energy
band for a lattice depth $u=E_{\rm R}$. Beyond being obviously zero
for $p_1=p_2$ the velocity $v$ also vanishes for $q>0$ along nontrivial
curves where the single particle group velocities are equal.
}
\label{vrelfig}
\end{figure}

In the presence of a lattice Wigner threshold laws describe elastic
and inelastic processes when the initial or final state relative group
velocity approaches zero. This naturally generalizes the notion of
Wigner laws for free scattering, which is expressed in terms of momenta
or equivalently in terms of velocities. In a periodic lattice however,
unlike in free space, a zero group velocity does not only occur for $q
\to 0$; see Fig.~(\ref{vrelfig}) for the fundamental band case. 
Threshold laws will therefore also apply at locations in the
Brillouin zone where the group velocities of the two molecules are close,
under the condition that they must be expressed in terms of $v$ rather than $q$.
In fact, the Wigner laws essentially arise from the density of energy
states for the relative motion $\rho_\epsilon$, and in a lattice 
$\rho_\epsilon \propto (\partial E / \partial q )^{-1} = v^{-1}$.

With this proviso, threshold laws can be expressed on the same footing
as {\it e.g.} in Ref.~\cite{2011-AS-JPB-235201} with the scaling behavior typical of
dimension one. For bosons, the elastic collision rate ${\cal K}^{\rm
el,B} \sim v$ and both the {\it umklapp} and the reactive probability vanish
with the group velocity of the incoming particles as $P^{\rm r,u} \sim
v$. Moreover, if the relative group velocity $v^\prime$ of the products
of a superelastic {\it umklapp} collisions is small, the probability for the
{\it umklapp} process behaves as $P^{\rm u} \sim v^\prime$. The $P^{\rm u}$
probability drops therefore continuously to zero as the exit channels
for the superelastic process become energetically closed. 

The Wigner laws valid for bosons also hold for identical particles of
fermionic nature, with the exception of the elastic collision rate that
vanishes for $q \to 0$ as ${\cal K}^{\rm el,F} \sim v^3$ for particles
in the same energy band. Note that the reciprocal-space point of
coordinates $(q=k_{\rm L}/2,Q)$ can be brought into $(q=0,Q+k_{\rm L})$
by a reciprocal lattice vector translation of $k_{\rm L}$ parallel to
the $p_2$ axis.  As a consequence, the Wigner laws for $q \to0$ and $q
\to \pm k_{\rm L}/2$ are the same.

Finally, the Wigner laws imply that for bosons the quantity $g^{\rm eff}$
is finite at the lower or upper limit of the energy band occurring at $q=0$.  
If the energy band limit occurs at nonzero values of $q$ the $g^{\rm eff}$
vanishes. Similarly, for fermionic molecules the momentum dependent scattering lengths $a^{\rm
F}(q,Q)$ tends to a finite quantity if the band edge occurs at $q = 0$
and vanishes otherwise.

\begin{figure}[t!]
\centering
\includegraphics[width=1.0\columnwidth]{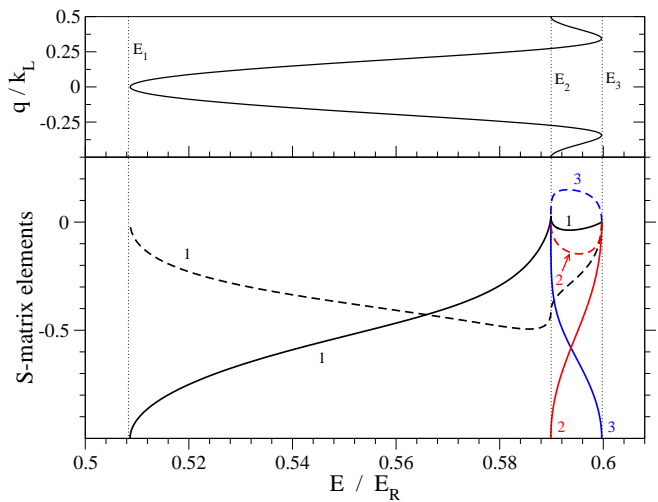}
\caption{
Upper panel : energy dispersion relation for two particles in a lattice
of depth $u=0.5~  E_{\rm R}$ and $Q=0.4~ k_{\rm L}$. Dotted lines denote
energy thresholds for collision channels (see text). Lower panel : real (full line)
and imaginary part (dashed line) of scattering matrix elements. Labels $1,2$ and $3$
denote the elements $S_{11}$, $S_{22}$, and $S_{12}$ defined in the text, respectively.
}
\label{figW}
\end{figure}

Les us examine in more detail the threshold behavior of the bosonic
scattering matrix elements. To this aim, we fix a specific value of
$Q=0.4 ~ k_{\rm L}$  such that energy band limits of different nature
$E_i$ ($i=1,2,3$) exist in the dispersion relation; See upper panel of
Fig.~(\ref{figW}). If $E<E_1$ no open channels (propagating waves)
exist. As energy increases above $E_1$ one symmetrized collision channel
becomes available. As shown in the lower panel of Fig.~(\ref{figW})
$E\to E_1^+$ the unique $S$-matrix element $S_{11} \to -1$.  In terms
of distinguishable particles this behavior would amount to reflection
with unit probability.

At the second threshold $E_2$ an additional channel with $q=k_{\rm
L}/2$ becomes energetically open. The matrix element $S_{11}$ is therein
continuous with discontinuous derivative, the new elastic element $S_{22}
\to -1$ as $E\to E_2^+$, whereas the transition amplitude $S_{12}$
vanishes at threshold.  Finally, for $E \to E_3^-$ the two open
channel states tend to coalesce and the $S$-matrix tends to a finite
limit with elements $S_{11}=S_{22}=0$ and $S_{12}=-1$. This result is
somehow expected, since exactly at threshold $E_3$ there is only one
quantum state and the rates for elastic and {\it umklapp} scattering
must therefore become equal there.

The fermionic scattering matrix elements simply behave as $S^{\rm F}=-S^{\rm B}$.

\subsection{One and two-channel models}
 \label{smodels}

Is is instructive to display at first the explicit structure of
the scattering matrix within a minimal model with only open channels
therefore neglecting the influence of the closed channels. Let us first
consider the bosonic case. Calculation in the presence of multiple open
channels is more easily performed using standing wave solutions defined
as ${\boldsymbol \eta}_1=\left( {\boldsymbol \eta}_++{\boldsymbol
\eta}_- \right)/2$ and ${\boldsymbol \eta}_2=-i \left({\boldsymbol
\eta}_+-{\boldsymbol \eta}_- \right)/2$.  Such asymptotic reference
function are real by virtue of the symmetry property ${\boldsymbol
\eta}_-^*={\boldsymbol \eta}_+$ below Eq.~\eqref{eqqr}.  Derivative
matrices ${\boldsymbol \xi}_1$ and ${\boldsymbol \xi}_2$ are likewise
represented in terms of ${\boldsymbol \xi}_+$ and ${\boldsymbol \xi}_-$.

By replacing standing waves for propagating waves, the asymptotic 
form \eqref{asyfunmat} defines the reactance
matrix $\mathbf K$. System of equations \eqref{zseq} becomes
\be
\label{sysK}
{\boldsymbol \xi}_2(0^+) 
  {\mathbf K}  =  {\mathbf Z} {\boldsymbol \eta}_1(0)  .
\ee
Right multiplication of the system \eqref{sysK} by $ {\boldsymbol
\eta}_{1}^t(0)$ gives the solution ${\mathbf K}= ({\boldsymbol \eta}_1^t
{\boldsymbol \xi}_2)^{-1}({\boldsymbol  \eta}_1^t {\mathbf Z} {\boldsymbol
\eta}_1)$, where the $0^+$ argument has been omitted.

Observing that ${\boldsymbol \eta}_2 (0)$ vanishes, the matrix
${\boldsymbol \eta}_1^t(0)  {\boldsymbol \xi}_2 (0^+) $
is identified as the value at the origin of the Wronskian
 \be 
\label{wronsk}
\mathbf W = {\boldsymbol
\eta}_1^t(z)  {\boldsymbol \xi}_2 (z) - {\boldsymbol \xi}_1^t(z)
{\boldsymbol \eta}_2 (z)
\ee
of linearly independent solution matrices of the Schr{\"o}dinger equation
for a given $Q$. Since the Wronskian is a constant matrix,
integrating both sides of Eq.~\eqref{wronsk} with respect to $z$ and
using Eq.~(E.6) in the appendix of Ref.~\cite{Ashcroft} to recover the group velocity
as well as the symplectic orthogonality condition below Eq.~\eqref{eqqr}
one obtains that ${\mathbf W}=a^{-1} {\mathbf 1}$. Had we used momentum-
rather than energy-normalized reference functions, the Wronskian would
have been proportional to a diagonal matrix with entries the relative
velocities. When applied to closed channels Eq.~(E.6) 
defines a purely imaginary velocity that, while nonphysical, will play an important 
role in the following.

The scattering matrix is expressed in terms of $\mathbf K$ using the
Cayley transform ${\mathbf S}=({\mathbf 1}+i{\mathbf K})({\mathbf
1}-i{\mathbf K})^{-1}$. The present approach is equivalent to the
unitarized Born approximation, where the reactance matrix is computed in
the first order Born approximation and the scattering matrix obtained
through the Cayley transform is automatically unitary if interaction
is real.

To be definite, let us consider the situation where two channels
characterized by relative quasimomentum $q_{i}$ and corresponding velocity
$v_i \equiv v_\alpha(q_i,Q)$ ($i=1,2$) are energetically available in
two-particle band $\alpha$ for total quasimomentum $Q$.  Simple matrix
algebra gives
\begin{widetext}
\be
\label{gtb}
    S =
 \left(
\begin{array}{cc}
 \dfrac{(g^\alpha_{12})^2 -(g^\alpha_{11}+i \hbar v_1) (g^\alpha_{22}-i\hbar v_2) }{(g^\alpha_{11}-i\hbar  v_1) (g^\alpha_{22}-i\hbar v_2)-(g^\alpha_{12})^2  }  &
\dfrac{2i\hbar  g^\alpha_{12} \sqrt{v_1 v_2} }{(g^\alpha_{11}-i\hbar  v_1) (g^\alpha_{22}-i\hbar v_2)-(g^\alpha_{12})^2}  \\ [2ex]
\dfrac{2i\hbar  g^\alpha_{21} \sqrt{v_1 v_2} }{(g^\alpha_{11}-i\hbar  v_1) (g^\alpha_{22}-i\hbar v_2)-(g^\alpha_{12})^2}  &
  \dfrac{(g^\alpha_{12})^2 -(g^\alpha_{11}-i \hbar v_1) (g^\alpha_{22}+i\hbar v_2) }{(g^\alpha_{11}-i\hbar  v_1) (g^\alpha_{22}-i\hbar v_2)-(g^\alpha_{12})^2}  \\ [2ex]
\end{array}
\right) ,
\ee
\end{widetext}
where the dependence on velocities has been made explicit.
The renormalized coupling constants $g^\alpha_{ij}$ in \eqref{gtb} are defined as the entries
of the symmetric matrix
\be
\label{gborn}
     {\mathbf g}^\alpha = g a  {\boldsymbol \eta}_1^t(0) {\boldsymbol \eta}_1(0) ,
\ee
in which the reference functions $\boldsymbol \eta$ must now be taken as
momentum normalized. In strong lattices the coupling constant scales with
the lattice depth as $g^\alpha_{ij} \sim u^{1/4}$ \cite{2001-PP-PRL-220401} whereas
$v$ decreases exponentially with $\sqrt{u}$ \cite{MPF-1989-PRB-546}.

Expression \eqref{gtb} explicitly shows for the particular case of two
open channels the Wigner laws stated in the previous section. Indeed,
for $E \to E_3^-$ one has $v_1 \simeq v_2 \to 0$, $ g^\alpha_{12}
\simeq g^\alpha_{11} \simeq g^\alpha_{22}$, implying $S_{12}\to -1$ and
$S_{ii}\to 0$. For $E\to E_2^+$ the velocity $v_1$ is finite whereas
$v_2 \to 0$. Crossing the threshold $E_2$ from above, the velocity
$v_2$ turns from purely real to imaginary giving rise to a singularity
in the energy derivative of the scattering matrix element $S_{11}$ (the
threshold singularity) which remains otherwise continuous. Below $E_2$
there is a sole physical matrix element $S_{11}$. Taking the limit $E
\to E_1^+$ ($v_1 \to 0$ with $v_2$ finite) one obtains $S_{11} \to -1$.

A simpler expression can be obtained in the single channel case,
where solution of the linear system \eqref{sysK} and Cayley transform
gives the unique scattering-matrix matrix element
\be
\label{gtb1}
   S=-\frac{g^\alpha_{11} +i \hbar v_1}{g^\alpha_{11} -i \hbar v_1} .
\ee
which evidently tends to $-1$ as $v_1$ vanishes.
The corresponding effective coupling constant is simply $g_\alpha^{\rm
eff}=g^\alpha_{11}$ and the reaction probability reads
\be
\label{1chPrEx}
P^{\rm r}_\alpha=- \frac{4 \hbar v_1 \Im{g^\alpha_{11}}}{(\Re{g^\alpha_{11}})^2+(\hbar v_1-\Im{g^\alpha_{11}})^2 }.
\ee
We recall that in general $\Im {g^\alpha_{ij}} < 0$ and that within the
universal model $\Re{g^\alpha_{11}}  = - \Im{g^\alpha_{11}} $. 
Analogous expressions for fermions are obtained by replacing the constant
$g$ with $g_{\rm map}$ and by changing the sign of $S$.  With these
modifications, one retrieves in particular the $K^{\rm el} \sim v^3$
law valid near the edge of an energy band $E_{mm}$ 
when the former occurs at $q=0$.

\subsection{Tight binding limit}

Single particle Bloch waves in energy band $n$ can in general be
represented in terms of Wannier states $w_n$ as $\phi^n_p (x)=\sum_s
e^{i s p a} w_n(x - sa) $.  This representation is most useful in the
tight binding limit, where the Wannier states are localized in individual
lattice sites with little overlap from site to site.  In this case, the
renormalized coupling constant becomes essentially independent of momentum
(hence $\forall i,j$ we will pose $g_{ij}^\alpha \equiv {\bar g}^\alpha$)
and only depends on the considered band indices $(m,n) =\alpha$ for the
two particles.
Therefore, the renormalized constant for single particle energy-bands takes the value
\be
 \label{gmodel}
   {\bar g}^\alpha =  a g \int_{-a/2}^{a/2}  w_m^2(z) w_n^2(z)  {\ud}z .
\ee 
By convenience we fix below the zero of energy at $p_{1,2}=k_{\rm L}/4$
in single-particle bands.

Let us consider first a one-channel model situation where two atoms are
in the same band.  In order to express $S$ in Eq.~\eqref{gtb1} in terms
of $Q$ and $E$ as independent variables, one needs to solve for $q$
the trigonometric equation
\be
    E_{mm}(q,Q)=  - 2J_m  \cos(a Q/2) \cos(a q) = E
\ee
and to insert the result in the expression for the group velocity
\be
   \hbar v=  2J_m a  \cos(a Q/2) \sin(a q ) .
\ee
With the position $E_Q \equiv E_\alpha(0,Q)$, simple algebra gives the relative group velocity
\be
   \hbar  v_1=  a \sqrt{E^2_Q - E^2}  .
 \label{velmodel}
\ee
Inserting Eq.~\eqref{gmodel} and \eqref{velmodel} into
\eqref{gtb1} one obtains the scattering matrix element 
\be
   S =   \frac{  \sqrt{ E^2_Q -E^2}-i {\bar g}/a }
   {  \sqrt{ E^2_Q - E^2 }+i {\bar g}/a }
\label{1chBH}
\ee
The present result is equivalent to the one
of Ref.~\cite{2008-NN-PRA-023617} that was obtained directly
in the Wannier representation.

Similarly, if the atoms are in different energy bands, the two solutions
of the trigonometric equation
\bea
    E_{mn}(q,Q) &=&  -J_+  \cos(a Q/2) \cos(a q) \nonumber \\ 
               &+& J_- \sin(a Q/2) \sin(a q)= E
\eea
with $J_+= J_m +J_n$ and $J_-=J_n m- J_n$
inserted in the corresponding expression for group velocity,
\be
  \hbar v  =   J_+ a \cos(a Q/2) \sin(a q) + J_- a \sin(a Q/2) \cos(a q)
\ee
give two coincident solutions
\be
\hbar v_{1,2}= a  \sqrt{J_+^2 \cos^2(a Q/2) + J_-^2 \sin^2(a Q/2) - E^2} .
\ee
Such identity of $v_1$ and $v_2$ only holds for the tight binding dispersion
relation and not in weak lattices.
Defining ${\bar v} \equiv v_{1,2}$ and using Eq.~\eqref{gtb} one obtains the reaction
probability as
\be
\label{pr2ch}
 P^r_\alpha= 1 -|S_{11}|^2-|S_{12}|^2= - \frac{4 \hbar {\bar v} \Im{\bar g}}{4 (\Re {\bar g})^2 +(\hbar {\bar v}-2 \Im {\bar g})^2} .
\ee
Again, in the universal model the latter expression can be simplified using $\Re {\bar g} = - \Im {\bar g}$.

\section{Results}
 \label{sec3}

We now present our main numerical results. We consider physical parameters
for K-Rb molecules in an optical tube with transverse angular frequency
$\omega_\perp = 2\pi\times40$~KHz. The longitudinal lattice is produced by
a laser of wavelength $2 \pi / k  = 1064$~nm. Long-range intermolecular
interactions in the absence of polarizing electric fields are isotropic
and will be parametrized in terms of the van der Waals length ${\bar
a}=118~a_0$ \cite{SK-2010-NJP-073041}. 

\subsection{Collision probabilities and rates}
\begin{figure}[!hb]
\centerline{
\epsfig{file=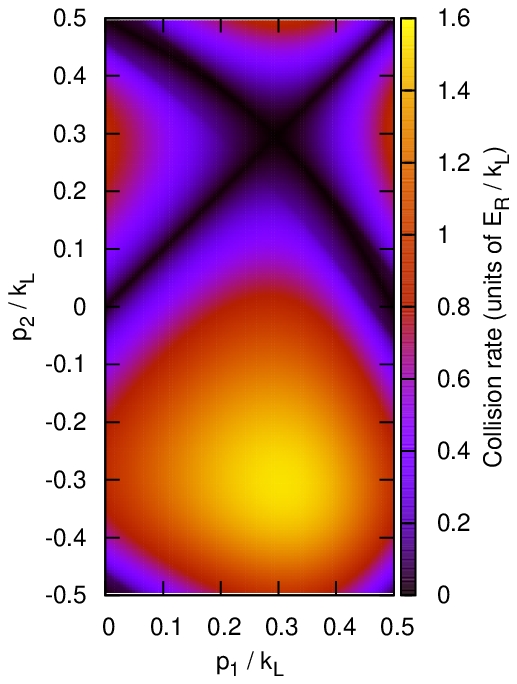,width=.5\columnwidth}
\epsfig{file=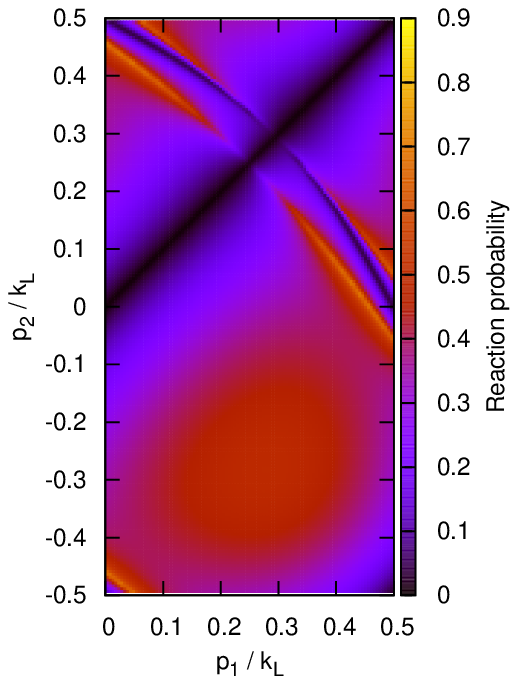,width=.5\columnwidth} }
\caption{
(color online) Elastic scattering rate (left panel) and reaction probability (right
panel) for collisions of bosonic molecules in a lattice
of depth $u=E_{\rm R}$. Both colliding molecules are in the fundamental
energy band.
}
\label{fig3} 
\end{figure}

Let us first consider bosonic molecules colliding in the fundamental
energy zone for a lattice of average strength $u=E_{\rm R}$.  The elastic
collision rate is shown in the left panel of Fig.~(\ref{fig3})
and it varies widely in the Brillouin zone. As expected it tends to be
maximum where $|v|$ is large and according to the Wigner laws it drops
to zero at the locations where $v \to 0$. The reaction probability
depicted in the right panel of the figure vanishes at the specific
locations where $v$ does. It also exhibits in the upper quadrant two
nontrivial maximum lines where $P^{\rm r} \sim 0.9$. These lines follow
closely the boundary separating the region where scattering is purely
elastic from the region where {\it umklapp} processes become allowed by
$E$ and $Q$ conservation. The maximum corresponds to a non-analyticity
cusp point or threshold singularity arising from the physics of channel
opening discussed in Sec.~II F; See also~\cite{1959-JRN-PR-1611} for a
comprehensive discussion.

\begin{figure}[hb!]
\centerline{
\epsfig{file=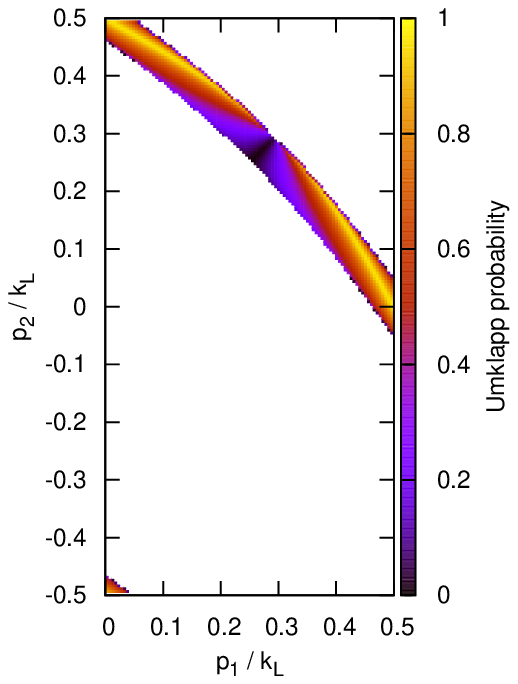,width=.5\columnwidth}
\epsfig{file=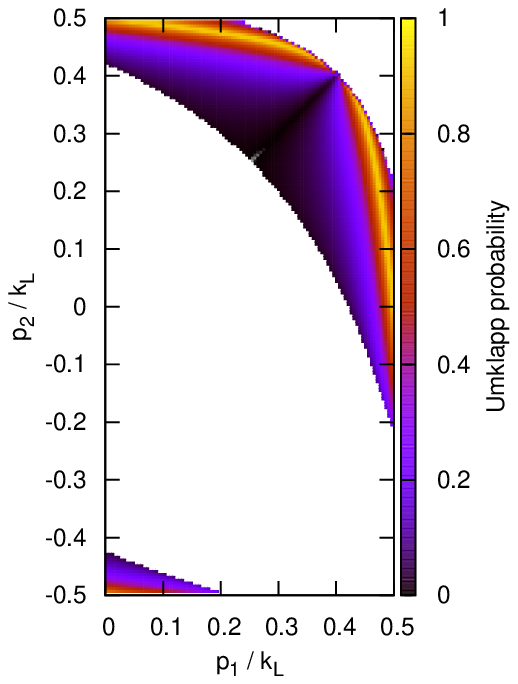,width=.5\columnwidth} }
\caption{
(color online) {\it Umklapp} collision probability for bosonic molecules
in a lattice of depth $u= E_{\rm R}$ (left panel) and $u=0.2~ E_{\rm R}$
(right panel). Both colliding molecules are in the fundamental energy band.  }
\label{fig4} 
\end{figure}

Right panel of Fig.~\ref{fig4} shows the probability of {\it umklapp} transitions. 
Note the region of the first Brillouin zone where
such processes are possible is quite narrow for $u=E_{\rm R}$. For
comparison, we show in the left panel the corresponding
results obtained for $u=0.2~ E_{\rm R}$, where the {\it umklapp} region
is significantly broader. In both cases, as per the Wigner laws the
probability drops to zero at the edges of the region, where the product
velocity $v'$ vanishes. The location near the middle of the region
where the probability reaches unity corresponds to the upper limit of
the relative-motion energy band occurring at $q \neq 0$. Along this line
there is coalescence of two quantum states and as discussed in Sec.~II E
according to the Wigner threshold laws $S_{12}\to -1$.

\begin{figure}[!ht]
\centerline{
\epsfig{file=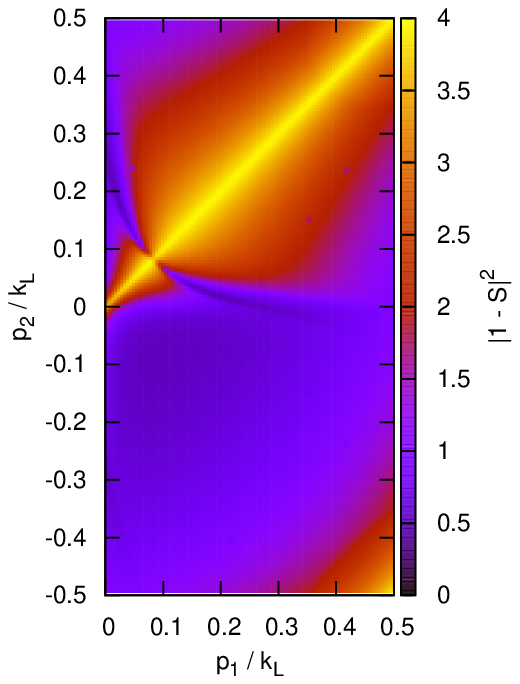,width=.5\columnwidth}
\epsfig{file=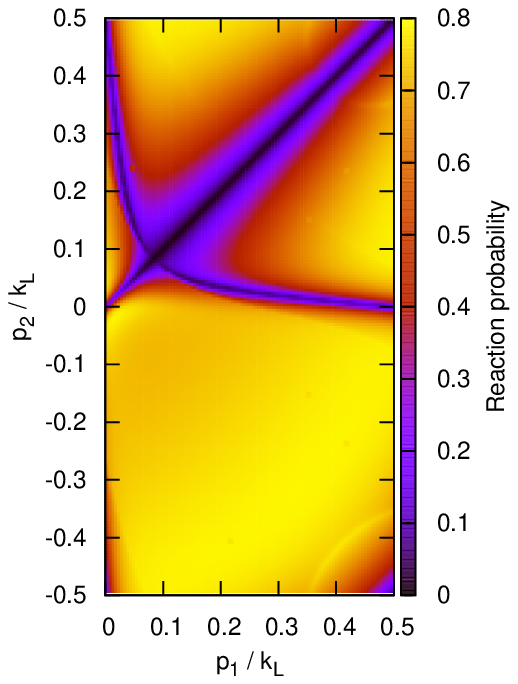,width=.5\columnwidth} }
\caption{
(color online) Elastic scattering factor (left panel) and reaction probability (right
panel) for collisions of bosonic molecules in a lattice
of depth $u=E_{\rm R}$. Both colliding molecules are in the first excited
energy band.
}
\label{fig5} 
\end{figure}

Fig.~\ref{fig5} presents our results for two particles in the first excited
energy band. For $u= E_{\rm R}$ the bandwidth is relatively large. As
shown in the right panel of the figure this increased mobility results
in the reaction probability being large in most of the Brillouin zone.
The elastic rate (not shown) presents a structure in the Brillouin zone
similar to the reactive one, mostly determined by the velocity factor
in definition \eqref{kel}.  At variance with the case of the fundamental
band we prefer to depict in the left panel the elastic scattering factor $|1-S(q Q \alpha  \gets q Q \alpha)|^2$.
Such factor reaches its maximum for $q \to 0$
($p_1 \to p_2$) where the elastic scattering matrix element $S \to -1$.
In most of the Brillouin zone both inter- and intra-band $\it umklapp$
processes are allowed with up to four channels open. Near the upper and
lower right corners of the Brillouin zone scattering is purely elastic.
Channel opening takes place along nontrivial lines where the scattering
quantities present nonanalytical cusp behaviors.  Closer inspection
shows for instance that the arched dark line in the left panel and the
two bright light lines near the upper and lower right corners are in
fact threshold features.

\subsection{Influence of the lattice depth}
Let us now study the possibility to use an optical lattice to control the
molecular reactivity. We fix example single particle quasi-momenta
$p_1=k_{\rm L}/4$ and $p_2=-p_1$ such that the reaction probability
tends to be large, as can be seen in the right panel of Fig.~(\ref{fig3}).

\begin{figure}[ht!]
\centering
\includegraphics[width=0.9\columnwidth]{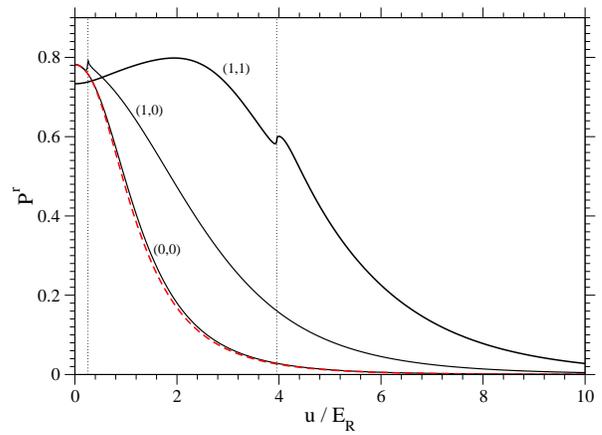}
\caption{
Reaction probability for two bosonic molecules with quasimomentum respectively equal
to $p_1 = k_{\rm L}/4$ and $p_2 = - k_{\rm L}/4$ as a function of the lattice
depth. Labels $(m,n)$ indicate the respective energy zone of each particle. 
Cusps in the $(1,0)$ and $(1,1)$ reaction probabilities correspond to collision thresholds
occurring at the values of $u$ marked by vertical dotted lines.
The dashed curve represents the one-channel approximation of Eq.~\eqref{1chPrEx}.
}
\label{figrub} 
\end{figure}

For particles in the fundamental band, Fig.~(\ref{figrub}) shows a
monotonic decrease of the reaction probability. The drop of $P^{\rm
r}$ with $u$ is slower for $u \lesssim E_{\rm R}$ and 
exponential for large $u$, as dictated by the velocity factor in the
numerator of Eq.~\eqref{1chPrEx}. 
The figure also shows a good agreement between the exact numerical result and the single-channel approximation 
of Eq.~\eqref{1chPrEx} with
interaction and velocity parameters calculated numerically.
The behavior of $P^{\rm r}_{\alpha}$
for $\alpha =(10)$ follows a similar pattern, with the exception of
a cusp point for $u \simeq 0.25~ E_{\rm R} $ corresponding to 
a collision channel closing as neighboring
energy bands separate. The value of $P^{\rm r}$ is about one order of
magnitude larger as compared to the case where the particles are in the
fundamental band.  For the case of two particles in the first excited
band, the reaction probability presents a non-monotonic behavior before
dropping exponentially.

\begin{figure}[ht!]
\centering
\includegraphics[width=0.9\columnwidth]{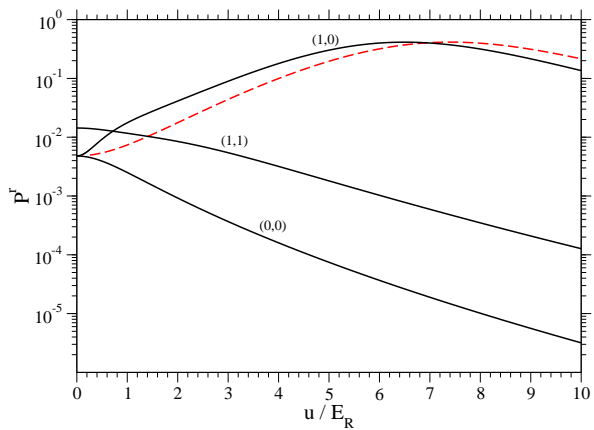}
\caption{
Same as Fig.~(\ref{figrub}) but for fermions. 
Note the dramatic increase of the reaction probability for particles
in $(1,0)$ bands.
Unlike the bosonic case, threshold singularities are not visible
on this scale.
The dashed curve represents the reaction probability obtained from the two-channel scattering
matrix of Eq.~\eqref{1chPrEx} with interaction and velocity parameters calculated numerically. 
}
\label{figruf} 
\end{figure}

As it can be seen in Fig.~(\ref{figruf}) the reaction probability
for fermionic molecules belonging to the same single-particle energy
band is naturally small in the absence of the lattice and is rapidly
decreasing with $u$. 
The situation is drastically different for particles belonging to the
fundamental and the first band that show a remarkably high reactivity
$P^{\rm r} \simeq 0.4$ for lattices of few recoils. 
As shown in the figure the two-channel model Eq.~\eqref{gtb} with $g$ replaced by $g_{\rm map}$ 
and the overall minus sign
needed for fermions is sufficient
to explain with sufficient accuracy this peculiar result. 
Further insight can be gained by noticing
that for weak lattices the CM quasimomentum can be identified with the
true CM momentum. Since the latter is conserved under the intermolecular interaction, 
for $u \ll E_{\rm R}$ one has $g_{12}^\alpha \approx 0$. 
Moreover, our pseudopotential is assumed independent of collision energy,
implying that to a good approximation $g_{11}^\alpha \approx g_{22}^\alpha$. In the
tight-binding limit one must obtain $g_{11}^\alpha \approx g_{22}^\alpha
\approx g_{12}^\alpha$. As remarked below Eq.~\eqref{map2} $g_{\rm map}$ is
large for fermions, magnitude is on the order of $10^2 \hbar v$ for our
physical parameters when $u$ is small. The reaction probability obtained
from the two-channel model 
can therefore be developed to first order in the
small parameter $v$. Simple algebra gives
\be
P^{\rm r}= \frac{2\hbar v  \Re g_{11}^\alpha }{(\Re g_{11}^\alpha)^2-(\Re g_{12}^\alpha)^2}  .
 \ee
The steep rise of $P^{\rm R}$ arises from the denominator exponentially vanishing with $u$.
In the tight binding limit the reaction probability is ruled by Eq.~\eqref{pr2ch} which 
gives the expected exponential drop of $P^{\rm r}$ with $v$.
Interpolation between these two trends results in the maximum observed in the exact numerical calculation.

\subsection{Effective coupling constant and CM motion}
\begin{figure}[ht!]
\centering
\includegraphics[width=0.9\columnwidth]{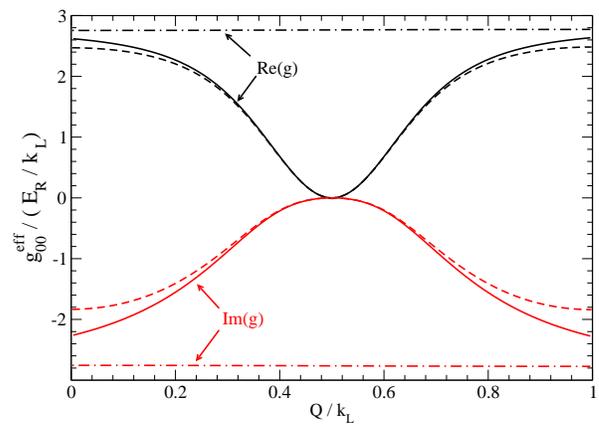}
\caption{
The real and imaginary parts of the effective coupling constant for
a lattice depth $u=5.0~E_{\rm R}$ computed in the $q\to 0$ limit as a function of the
total quasimomentum $Q$. Numerical results (full line) are compared with
single channel (dash-dot lines) and two-channel (dashed lines) models (see text).
}
\label{figvsQ} 
\end{figure}

The presence of a lattice breaks up Galilean invariance of flat
space. Scattering quantities in a lattice will therefor in general
depend on the value of $Q$ or, more precisely, on the CM group velocity
$v_{\rm CM}$. Consider an ultracold gas such that in the
center-of-mass frame the atom quasimomenta $|p_i| \ll k_{\rm L}$. In the
laboratory frame $Q$ does not need to be small, a situation that can be
experimentally realized by creating an ultracold gas at rest
and then adiabatically accelerating or tilting the lattice~\cite{2005-LDS-PRA-013603}.
We consider here the case of gas with $|q \ll k_{\rm L}|$ and focus on a
relatively deep lattice $u=5~E_{\rm R}$. 
The numerically calculated $g^{\rm eff}$ is shown in Fig.~(\ref{figvsQ}) for bosonic
molecules colliding in the fundamental band. Most striking
feature is the drop of both real and imaginary parts of the effective coupling
constant for $Q\simeq k_{\rm L}/2$ ($p_{1,2} \simeq k_{\rm L}/4$).

In fact, Eq.\eqref{1chBH} results from the scattering solution of a tight
binding Hamiltonian with only one symmetrized state of relative motion,
equivalent to the Bose-Hubbard model with two particles. This expression
predicts an effective coupling constant $g_\alpha^{\rm eff}= \bar
g^\alpha$. The momentum dependence of functions $u_q$ in Eq.~\eqref{gborn}
and thus of $\bar g$ becomes negligible in the strong lattice limit
and the dispersion relation in lattices of depth beyond few recoils is
also very accurately represented by the tight binding form. The strong
dependence of $\bar g$ on quasimomentum depicted in Fig.~(\ref{figvsQ}) is
therefore unexpected.

For the considered lattice depth the tight binding approximation holds
with accuracy and {\it umklapp} transitions are forbidden for any $Q$.
However, one exception exists and is the special value $Q=k_{\rm L}/2$ at
which according to Eq.~\eqref{Enm} with $n=m$ the relative energy band
becomes completely flat. This nonphysical feature stems from the strict
($u \to \infty$) tight binding approximation but in practice the relative
motion energy presents for $Q\simeq k_{\rm L}/2$ extrema analogous to the
dashed curves shown in Fig.~(\ref{fig1}) for weaker lattices. In this
situation a pair of symmetrized open channels with real $q$ exists. As
$Q$ moves away from $k_{\rm L}/2$ one of the two open channels becomes
energetically closed ($q$ becomes imaginary) and thus inaccessible to
real transitions. However, virtual transitions to such closed channel
have a profound influence on the collision.

In order to study this effect on a quantitative ground, let us consider the
two-channel scattering matrix Eq.~\eqref{gtb} for one open channel and a
second channel which can now be either open or become closed with purely
imaginary velocity $v_2 \equiv i {\tilde v}_2$. In the latter case,
the relevant $S$-matrix element becomes
\be
S_{11}=  \dfrac{(g^\alpha_{12})^2 -(g^\alpha_{11}+i \hbar v_1)
(g^\alpha_{22}+ \hbar {\tilde v}_2) }{(g^\alpha_{11}-i\hbar
v_1) (g^\alpha_{22}+ \hbar {\tilde v}_2)-(g^\alpha_{12})^2 }
\ee
which through Eq.~\eqref{geff} gives an effective coupling constant 
\be
\label{geff2ch}
g_\alpha^{\rm eff} = \frac{g^\alpha_{11}(g^\alpha_{22}+ \hbar {\tilde
v}_2)-(g^\alpha_{12})^2 }{g^\alpha_{22}+ \hbar {\tilde v}_2}.
\ee

\begin{figure}[hb!]
\centering
\includegraphics[width=0.9\columnwidth]{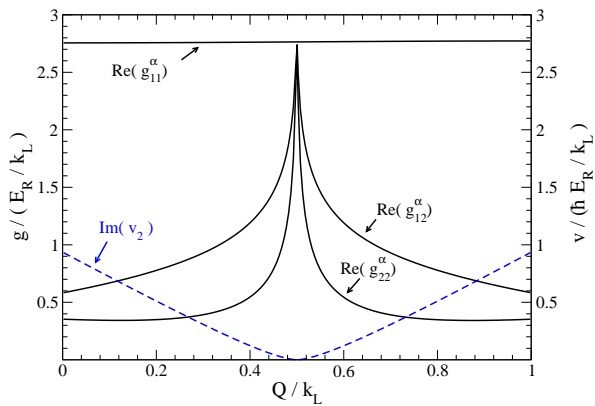}
\caption{
Evolution with $Q$ of the two-channel model parameters in Eq.~\eqref{gtb}
for a lattice depth $u=5~E_{\rm R}$. The vanishing of the velocity parameter $v_2$ 
near $Q=k_{\rm L}/2$ has a strong influence on the collision (see text).
The imaginary part of the coupling constant is not shown since for universal
reactions one has strictly $\Im g = - \Re g$.
}
\label{gijvfig} 
\end{figure}

The behavior of the parameters factoring in the former equation is represented in
Fig.~(\ref{gijvfig}), which shows namely the quantities $\Re g^\alpha_{ij}
= -\Im g^\alpha_{ij}$ and ${\tilde v}_2 =\Im v_2$ as a function of $Q$. On one
side, one may observe that as predicted by the tight binding approximation
the dependence of $g^\alpha_{11}$ on the particle quasi-momenta is
extremely weak. One the other side, the open-closed $g^\alpha_{12}$ and
closed-closed $g^\alpha_{22}$ channel couplings involve closed-channel
wavefunctions that cannot be represented in terms of localized Wannier
functions. As a consequence, both $g^\alpha_{12}$ and $g^\alpha_{22}$
show significant dependence on $Q$ and only tend to $g^\alpha_{11}$
for $Q$ near $k_{\rm L}/2$, where the closed and the open channels coalesce.
The velocity $v_2$ will be real and small in the tiny
region of $Q\approx k_{\rm L}/2$ where two open channel exists. Moving
away from this region $v_2$ first vanishes then evolves continously
into a purely imaginary quantity.

The single-channel result $g_\alpha^{\rm eff} = g^\alpha_{11}$ is only
retrieved in the case of weak interactions $| g^\alpha_{22} | \ll \hbar
|{\tilde v}_2 |$ and $|g^\alpha_{12}|^2 \ll \hbar | {\tilde v}_2 g^\alpha_{11}
|$. These conditions will however never be satisfied for collisions of
atoms with $q\approx 0$ and $Q\approx k_{\rm L}/2$ since ${\tilde v}_2$
becomes then arbitrarily small. In this case the effect of the closed
channel must be taken into account and Eq.~\eqref{geff2ch} predicts
indeed the noninteracting behavior with $g_\alpha^{\rm eff} \simeq 0$
observed in the full numerical calculation of Fig.~(\ref{figvsQ}). Note incidentally that
the denominator of Eq.~\eqref{geff2ch} may vanish if $g_{22}^\alpha$ is real
({\it i.e.} for non reactive species) and equal to ${\tilde v}_2$,
leading to novel resonant behaviors that we plan to study in the near future.

\section{Conclusions and perspectives}

We have presented a computational algorithm for the calculation of two-body collision
properties in an infinite periodic 1D structure. Our computational approach presents distinct advantages
since it uses regular and irregular functions computed at fixed collision energy, therefore avoiding
a numerically more tedious spectral expansion of the Green's operator \cite{2006-MW-PRL-012707}.
The model allows us to assess quantitatively the expected effect of the lattice in suppressing the
reactive collision rates. We also show that for fermionic molecules in different bands the lattice can have the
counterintuitive effect of strongly enhancing the reactive processes before suppressing them. 
A two-channel approach stresses the role of Bloch waves with complex relative quasimomentum in renormalizing 
the coupling constant for particles in the fundamental band even when the lattice is strong.

In perspective it will be interesting to study lattice-induced resonances in non-reactive systems and,
in the spirit of what has been done in three dimensions in Ref.~\cite{2010-HPB-PRL-090402}, to predict the
Bose-Hubbard parameters in 1D systems based on an accurate microscopic model.
Finally, if a fully quantitative model will be needed in order to compare with experiments, the present reference functions can be
used in standard 3D close-coupled models to extract efficiently the scattering observables.

\begin{acknowledgments}
We are indebted to E. Tiesinga and C. J. Williams for contributions to
the methodological developments presented in this work. We would like to
thank Z. Idziaszek for useful discussions and T. Kristensen for comments
on the manuscript. This work was supported by the Agence Nationale de
la Recherche (Contract No. ANR-12-BS04-0020-01).

\end{acknowledgments}

\end{document}